\documentclass[nato]{crckbked}
\numreferences
\usepackage{graphicx}
\usepackage{bm}

\usepackage{amstext}

\def\coleft{[}
\def\ccright{]}

\def\idline{} 
\makeindex
\begin{document}
\begin{article}

\begin{opening}
\title{Transport properties of correlated electrons in high dimensions}

\author{N. \surname{Bl\"{u}mer}\email{Nils.Bluemer@uni-mainz.de}}
\author{P.~G.~J. \surname{van Dongen\-}\email{Peter.vanDongen@uni-mainz.de}}
\institute{Institut f\"{u}r Physik, Johannes Gutenberg-Universit\"{a}t,\\
55099 Mainz, Germany}

\runningauthor{N.\ Bl\"{u}mer, P.\ G.\ J.\ van Dongen}

\begin{abstract}
  We develop a new general algorithm for finding a regular tight-binding 
  lattice Hamiltonian in infinite dimensions
  for an arbitrary given shape of the density of states (DOS). The
  availability of such an algorithm is essential for the investigation
  of broken-symmetry phases of interacting electron systems and for the
  computation of transport properties within the dynamical mean-field
  theory (DMFT). The algorithm enables us to calculate the optical
  conductivity fully consistently on a regular lattice, e.g., for the
  semi-elliptical (Bethe) DOS. 
  We discuss the relevant $f$-sum rule and present numerical results
  obtained using quantum Monte Carlo techniques.
\end{abstract}
\end{opening}


\section{Introduction}\label{sec:intro}
Microscopic studies of strongly correlated electron systems require
methods which take the Coulomb repulsion between electrons explicitly into
account. Many features of such systems can be modeled using the single-band
Hubbard model \cite{Hubbardetc}
\begin{equation}\label{eq:Hubbard}
\hat{H}=\hat{H}_0 + \hat{H}_\text{int} = \sum_{ij,\sigma} t_{ij} \;
\hat{c}_{{\bm{R}}_i\sigma}^\dagger \hat{c}_{{\bm{R}}_j\sigma}^{{\phantom{\dagger}}}
+ U \sum_{i} \hat{n}_{{\bm{R}}_i\uparrow} \hat{n}_{{\bm{R}}_i\downarrow} \,,
\end{equation}
where the operators $\hat{c}_{{\bm{R}}_i\sigma}^\dagger$ and
$\hat{c}_{{\bm{R}}_i\sigma}^{\phantom{\dagger}}$ create and destroy electrons of spin
$\sigma$ on site ${\bm{R}}_i$, respectively;
$\hat{n}_{{\bm{R}}_i\sigma}$ measures the corresponding occupancy. A
general nonperturbative treatment of this model is only possible in
the limit of infinite dimensionality where the dynamical
mean-field theory becomes exact: due to a local self-energy the model
reduces for $d\to\infty$ to a single impurity Anderson model plus a 
self-consistency
equation. For homogeneous phases, local properties then only depend on
the lattice via the noninteracting DOS 
$\rho(\epsilon):=\frac{1}{N}\sum_{\bm{k}} \delta(\epsilon-\epsilon_{\bm{k}})$.  
In the simplest case of
uniform nearest-neighbor hopping on a hypercubic (hc) lattice, the
dispersion reads 
$\epsilon_{\bm{k}}^{\text{hc}}=-2t \sum_{\alpha=1}^d \cos(k_\alpha)$
which for the proper scaling $t=t^*/\sqrt{2d}$ leads to a Gaussian DOS
$\rho^{\text{hc}}(\epsilon)=\exp[-\epsilon^2/(2{t^*}^2)]/(\sqrt{2\pi}t^*)$.

Vertex corrections to the optical conductivity $\sigma(\omega)$ vanish
\cite{Khurana90a} in the limit $d\to\infty$ so that it may be
expressed in the isotropic case as \cite{Pruschke93a}
\begin{equation}\label{eq:opt_dmft}
\sigma(\omega) = \sigma_0\!
\int_{-\infty}^\infty \!\!\mathrm{d}\epsilon\, \tilde{\rho}(\epsilon)
\int_{-\infty}^\infty \!\!\mathrm{d}\omega'\,
A_\epsilon(\omega') A_\epsilon(\omega'+\omega)
\frac{n_{\text{f}}(\omega')-n_{\text{f}}(\omega+\omega')}{\omega}\,.
\end{equation}
Here, $A_\epsilon(\omega)$ is the ``momentum'' dependent spectral
function, $n_{\text{f}}(\omega')$ is the Fermi function, 
$\sigma_0= \frac{2\pi e^2}{\hbar^2}\frac{N}{V}$, and 
\begin{equation}\label{eq:rhot}
\tilde{\rho}(\epsilon) := \frac{1}{N} \sum_{\bm{k}} |{\bm{v}}_{{\bm{k}}}|^2 \delta(\epsilon-\epsilon_{\bm{k}})
=: \langle |{\bm{v}}_{{\bm{k}}}|^2\rangle(\epsilon)\,\rho(\epsilon)\,.
\end{equation}
Note that the frequency- and interaction-dependent part in (\ref{eq:opt_dmft})
depends on the lattice only via the DOS while the explicitly lattice-dependent
part $\tilde{\rho}(\epsilon)$ is universal, i.e., independent of interaction $U$, filling,
and temperature.
In the hypercubic case, the momentum dependence of the Fermi velocity
${\bm{v}}_{{\bm{k}}}$ becomes irrelevant in (\ref{eq:rhot}) for $d\to\infty$; for unit hopping and
lattice spacing, one observes
$\tilde{\rho}^\text{hc}(\epsilon)=\rho^\text{hc}(\epsilon)$. As a
consequence, the optical $f$-sum is then proportional to the kinetic
energy: 
\begin{equation}\label{eq:fsum_ekin}
\int_0^\infty \!\mathrm{d}\omega \,\sigma(\omega)= - \frac{\sigma_0}{4}\langle
\hat{H_0}\rangle\,. 
\end{equation}

\begin{figure}
a)\hspace{0.31\textwidth} b)\hspace{0.28\textwidth} c)\vspace{-2ex}

\includegraphics[height=0.18\textheight,clip=true]{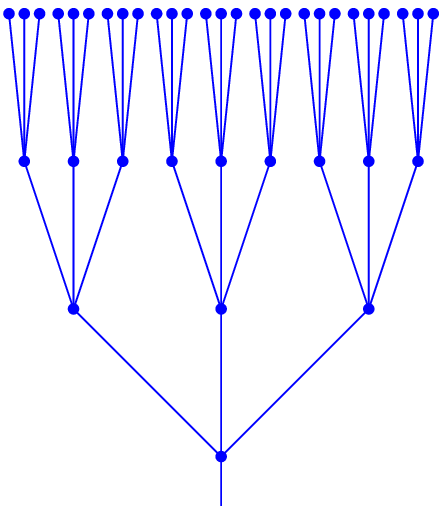}
\hspace{0.09\textwidth}
\includegraphics[height=0.2\textheight,clip=true]{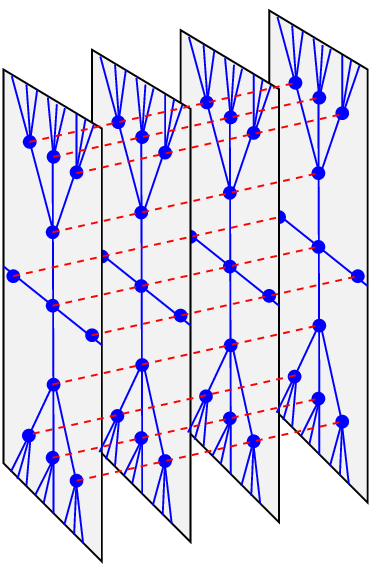}
\hspace{0.06\textwidth}
\includegraphics[height=0.2\textheight,clip=true]{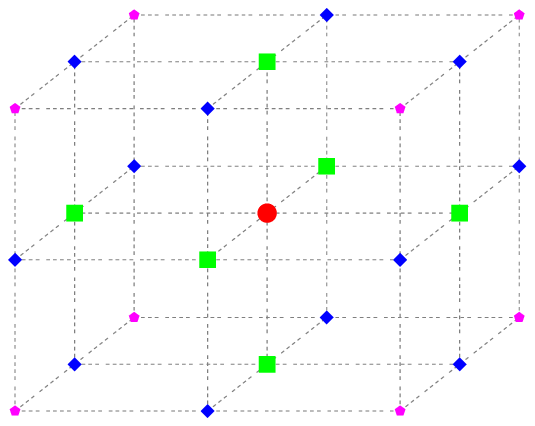}
\caption{Bethe lattice: {\bf a)} conventional tree level picture for
  coordination $Z=4$, {\bf b)} stacked Bethe lattice, {\bf c)}
  redefined Bethe lattice as a regular lattice with cubic symmetry:
  here, symbols denote equal hopping matrix elements relative to a
  fixed site (central circle).}
\label{fig:layout_bethe}
\end{figure}

However, the hc DOS is unbounded which is
hardly compatible with the single-band assumption. In fact, no regular
lattice model with sharp band edges in $d\to\infty$ could be
constructed so far. In this situation, many DMFT studies have focussed
on the so-called Bethe lattice which is not a regular lattice, but a
tree in the sense of graph theory as shown in Fig.~\ref{fig:layout_bethe}a.
The semi-elliptic DOS $\rho(\epsilon)=\sqrt{4-\epsilon^2}/(2\pi)$ of
this model (for $Z\to\infty$) fixes the local properties of the model;
transport, however, is a priori undefined.  A derivation of
$\sigma(\omega)$ directly for the Bethe tree (using the level-picture
Fig.~\ref{fig:layout_bethe}a) by Chung and Freericks \cite{Chung98a}
is still incomplete \cite{Bluemer02a}; up to a factor of 3, the same
expression $\tilde{\rho}(\epsilon)\propto
(4-\epsilon^2)\rho(\epsilon)$ was obtained \cite{Chattopadhyay00a} in
a heuristic scheme by enforcing the hc $f$-sum rule
(\ref{eq:fsum_ekin}). An alternative direct approach \cite{Stumpf99a}
fails to describe the coherent transport expected in the metallic regime.

The local DMFT problem is unchanged when a finite number of hopping
bonds per site are added.  Therefore, the periodically stacked Bethe
lattice (Fig.~\ref{fig:layout_bethe}b) is still a Bethe lattice
in the DMFT sense; potentially coherent transport is then
well-defined (only) in stacking direction \cite{Uhrig93b}. A semi-elliptic
DOS also results from fully disordered hopping on lattices of
arbitrary topology; in this case, $\sigma(\omega)$ is incoherent
\cite{Dobrosavljevic93a}. 

We will in the following construct and
evaluate a new definition for $\sigma(\omega)$ compatible with a semi-elliptic
DOS. This definition is unique by being based on a regular lattice as
illustrated in Fig.~\ref{fig:layout_bethe}c and by leading to an isotropic
conductivity which is coherent in the noninteracting limit.

\section{General Dispersion Method}
\label{sec:general}
We rewrite the translation-invariant noninteracting Hamiltonian,
\begin{equation}
\hat{H}_0 = \sum_{i,\sigma} \sum_{\bm{\tau}} t_{\bm{\tau}} \;
\hat{c}_{{\bm{R}}_i\sigma}^\dagger \hat{c}_{{\bm{R}}_{i}+{\bm{\tau}},\sigma}^{{\phantom{\dagger}}}
= \sum_{{\bm{k}},\sigma} \epsilon({\bm{k}})\, \hat{n}_{{\bm{k}}\sigma},
\end{equation}
where contributions to the dispersion may be classified by the
taxi-cab hopping distance 
$|\:\!\!|\bm{\tau}|\:\!\!|=\sum_{\alpha=1}^d |\tau_\alpha|$:
\begin{equation}\label{eq:energy_splitup}
\epsilon({\bm{k}})=\sum_{D=1}^\infty \epsilon_D({\bm{k}}),\qquad
\epsilon_D({\bm{k}})=\sum_{|\:\!\!|\bm{\tau}|\:\!\!|=D} t_{\bm{\tau}} e^{i{\bm{\tau}}\cdot{\bm{k}}}\,.
\end{equation}
In high dimensions, only vectors
of the form ${\bm{\tau}}=\sum_{i=1}^D {\bm{e}}_{\alpha_i}$ with pairwise different
directions $\alpha_i$ need to be considered. This follows from the fact
that the fraction of neglected vectors (with $|{\bm{\tau}}\cdot
{\bm{e}}_\alpha|>1$ for some direction $\alpha$) vanishes as $1/d$. 
Furthermore, the considered vectors are of minimal 
Euclidean length $|{\bm{\tau}}|$ hinting at maximal overlap, i.e., largest 
$|t_{\bm{\tau}}|$ for fixed taxi-cab distance $|\:\!\!|\bm{\tau}|\:\!\!|$ and fixed $D$.

By deriving a recursion relation for $\epsilon_D({\bm{k}})$ we have established
that
\begin{equation}\label{eq:general_dispersion}
\epsilon({\bm{k}}) \,=\, \sum_{D=1}^\infty \frac{t_D^*}{\sqrt{D!}}\; 
\text{He}_D(\epsilon^{\text{hc}}_{\bm{k}})\,=:\, 
{\cal F}(\epsilon^{\text{hc}}_{\bm{k}})\,.
\end{equation}
Using the orthogonality of the Hermite polynomials, one may express the
hopping matrix elements in terms of the transformation
function ${\cal F}(x)$:
\begin{equation}\label{eq:general_ts}
t_D^* = \frac{1}{\sqrt{2\pi D!}}\int_{-\infty}^\infty \!\mathrm{d}\epsilon \,{\cal F}(\epsilon) 
\,\text{He}_D(\epsilon)\, e^{-\epsilon^2/2}\,.
\end{equation}
Specializing on the case of a monotonic
transformation function ${\cal F}(x)$ (with derivative ${\cal F}'(x)$), we can write
\begin{equation}\label{eq:general_solution}
\rho(\epsilon)=
\frac{1}{{\cal F}'({\cal F}^{-1}(\epsilon))} \,\rho^{\text{hc}}({\cal F}^{-1}(\epsilon))
\end{equation}
which leads to
\begin{equation}\label{eq:general_comp_fges}
{\cal F}^{-1}(\epsilon)=\sqrt{2}\;\text{erf}^{-1}
\bigg(2\int_{-\infty}^\epsilon \!\mathrm{d}\epsilon' \rho(\epsilon')-1\bigg)\,.
\end{equation}
Furthermore, the Fermi velocity ${\bm{v}}_{\bm{k}}=\nabla \epsilon_{\bm{k}}$ can be computed:
\begin{equation}\label{eq:general_vksq}
{\bm{v}}_{\bm{k}} = {\cal F}'({\cal F}^{-1}(\epsilon)) {\bm{v}}_{\bm{k}}^{\text{hc}}
= \frac{\rho^{\text{hc}}(\sqrt{2}\,\text{erf}^{-1}(2\int_{-\infty}^\epsilon 
\!\mathrm{d}\epsilon' \rho(\epsilon')-1))}{\rho(\epsilon)} \, {\bm{v}}_{\bm{k}}^{\text{hc}}\,.
\end{equation}
A practical application of the general formalism proceeds
as follows:
\begin{enumerate}
\item compute ${\cal F}^{-1}(\epsilon)$ from arbitrary target DOS $\rho(\epsilon)$
using (\ref{eq:general_comp_fges})
\item invert function (numerically or analytically) to obtain ${\cal F}(\epsilon)$
\item evaluate transport properties, e.g., 
$\langle |{\bm{v}}_{\bm{k}}|^2\rangle(\epsilon)$ or $\tilde{\rho}(\epsilon)$
using (\ref{eq:general_vksq})
\item optionally determine microscopic model parameters $t_D^*$ using (\ref{eq:general_ts})
\end{enumerate}
The only choice inherent in this procedure beyond the usual
assumptions for large dimensions is contained in step 1 which by 
construction produces a monotonic transformation function ${\cal F}$. 
The optical $f$-sum rule reads
\begin{equation}\label{eq:fsum}
\displaystyle \int_0^\infty \!\mathrm{d}\omega \,\sigma_{xx}(\omega) 
= \displaystyle \frac{\sigma_0}{4d}\, \big\langle 
\frac{\tilde{\rho}'(\epsilon)}{\rho(\epsilon)}
\big\rangle
= \displaystyle\frac{\sigma_0}{4d}\, \left\langle\left[
{\cal F}''\big({\cal F}^{-1}(\epsilon)\big)-{\cal F}^{-1}(\epsilon){\cal F}'\big({\cal F}^{-1}(\epsilon)\big)
\right]\right\rangle\,.
\end{equation}
Here, the first equality follows from
(\ref{eq:opt_dmft}), i.e., is generally valid within the DMFT while
the second expression in terms of the transformation is specific to the
formalism developed within this section.

\paragraph{Example: Flat-band System}
One interesting limiting case of a monotonic transformation function
which can be treated analytically is the step function 
${\cal F}(x)=2\Theta(x)\!-\!1$
corresponding to a flat band DOS of the form
$\rho(\epsilon)=\big(\delta(\epsilon\!-\!1)+\delta(\epsilon\!+\!1)\big)/2$.  For this
case, the hopping matrix elements read (trivially, $t_{2n}^*=0$):
\begin{equation}\label{eq:flat_ts}
t_{2n+1}^* = (-1)^n \sqrt{\frac{2}{\pi}} \frac{(2n-1)!!}{\sqrt{(2n+1)!}}
\quad\stackrel{n\to\infty}{\longrightarrow}\quad
(-1)^n (\pi n)^{-3/4}\;.
\end{equation}
The asymptotic exponent $-3/4$ is only slightly smaller than the
threshold value $-1/2$ required for a finite variance 
$\int_{-\infty}^\infty d\epsilon \,\epsilon^2 \rho(\epsilon)=\sum_D
{t_D^*}^2$. For a rectangular model DOS, 
$t_D^*\sim 2^{-n} n^{-3/4}$ already decays exponentially fast. 


\section{Application to the ``Bethe'' semi-elliptic DOS}\label{sec:dos_bethe}

\begin{figure}
a)\hspace{0.48\textwidth} b)\vspace{-2ex}

\includegraphics[width=.49\textwidth,clip=true]{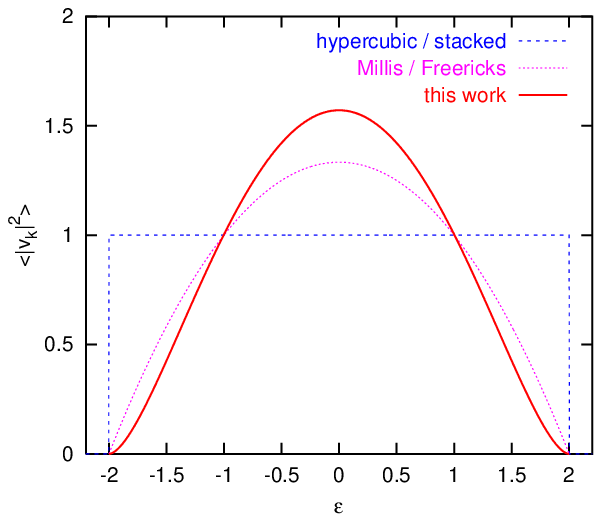}
\hfill \includegraphics[width=.49\textwidth,clip=true]{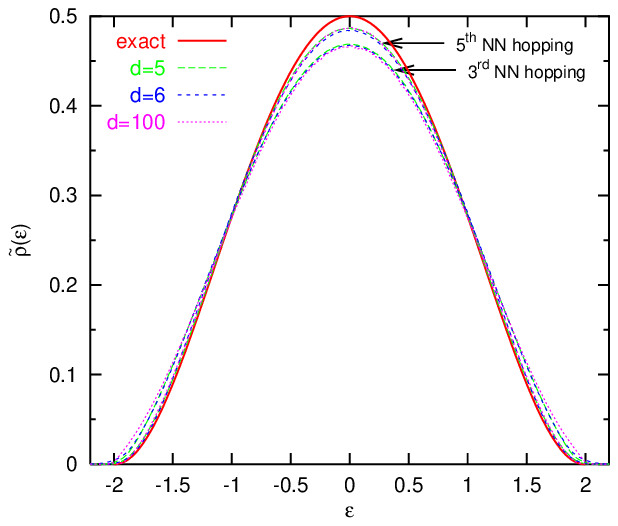}
\caption{{\bf a)} The average squared Fermi velocity 
  $\langle |{\bm{v}}_{\bm{k}}|^2\rangle(\epsilon)$ is constant for the
  hypercubic lattice (or for the $x$-component of a stacked lattice);
  in contrast it vanishes for the isotropic lattice defined in this
  work. For comparison, the form suggested by Millis is also shown.
  {\bf b)} Resulting function $\tilde{\rho}(\epsilon)$ of the full
  isotropic model (solid line) in comparison with truncated models
  ($D_{\max}=3$ or $D_{\max}=5$), evaluated in finite dimensions $5\le
  d\le 100$.}
\label{fig:vksq_rhot_bethe}
\end{figure}

In this section, we will apply the new formalism to the Bethe
semi-elliptic DOS in order to determine a corresponding tight-binding
Hamiltonian defined on the hypercubic lattice with the same local
properties as the Bethe lattice (with NN hopping) in the limit
$d\to\infty$.  From (\ref{eq:general_vksq}), we
derive the average squared Fermi velocity defined in (\ref{eq:rhot}) in
closed form:
\begin{equation}\label{eq:vksq_bethe_general}
\langle|{\bm{v}}_{\bm{k}}|^2\rangle(\epsilon)=\frac{2\pi}{4-\epsilon^2}
\exp\left[-2 \left(\!\text{erf}^{-1}\Big(
\frac{\epsilon\sqrt{1-\epsilon^2/4}+2 \arcsin(\epsilon/2)}{\pi}
\Big)\!\!\right)^{\!\!2}\right]\,.
\end{equation}
Here, we have used the fact that $\langle|{\bm{v}}_{\bm{k}}|^2\rangle(\epsilon)$
is effectively constant (and equals 1 for unit variance and lattice
spacing) in the hypercubic case.
The result (solid line in Fig.~\ref{fig:vksq_rhot_bethe}a)
 has all the qualitative
features expected for this observable in any finite dimension:
$\langle|{\bm{v}}_{\bm{k}}|^2\rangle(\epsilon)$ is maximal near the band center,
strongly reduced for large (absolute) energies and vanishes at the
band edges: states at a (noninteracting) band edge do not contribute
to transport. The violation of this principle in the stacked case
(dashed lines in Fig.~\ref{fig:vksq_rhot_bethe}a), which 
corresponds to an application
of the hc formalism to the Bethe DOS with $\langle|{\bm{v}}_{\bm{k}}|^2\rangle$
constant up to the band edges, is clearly pathological. Therefore, our
method has not only the merit of yielding isotropic transport, but also
of avoiding unphysical behavior.

In order to determine the microscopic model, we have to apply
(\ref{eq:general_ts}) to the numerically evaluated transformation function
${\cal F}$. Again, the scaled hopping
matrix elements fall off exponentially fast: only a fraction
$10^{-3}$ of the total energy variance arises from hopping amplitudes
beyond third nearest neighbors and only a fraction $10^{-6}$ results
from hopping beyond $9^\text{th}$-nearest neighbors. This result suggests
that properties of the model should be robust with respect to
truncation.  In fact, $\tilde{\rho}(\epsilon)$ (and consequently the definition of
$\sigma(\omega)$) hardly changes when hopping is cut off beyond 
$3^{\text{rd}}$ or
$5^{\text{th}}$ nearest neighbors, even when evaluated in finite
dimensions as seen in Fig.~\ref{fig:vksq_rhot_bethe}b. This behavior is very
general so that results for $\sigma(\omega)$ of a local theory in finite
dimensions will depend on $d$ predominantly via the
interacting DOS $A(\omega)$ and only very little via $\tilde{\rho}(\epsilon)$.

\begin{figure}
a)\hspace{0.48\textwidth} b)\vspace{-2ex}

  \includegraphics[width=0.49\textwidth,clip=true]{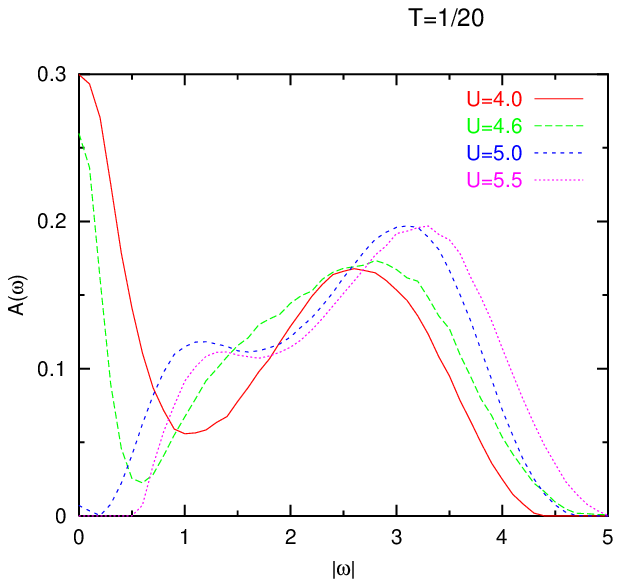}
  \hfill \includegraphics[width=0.49\textwidth,clip=true]{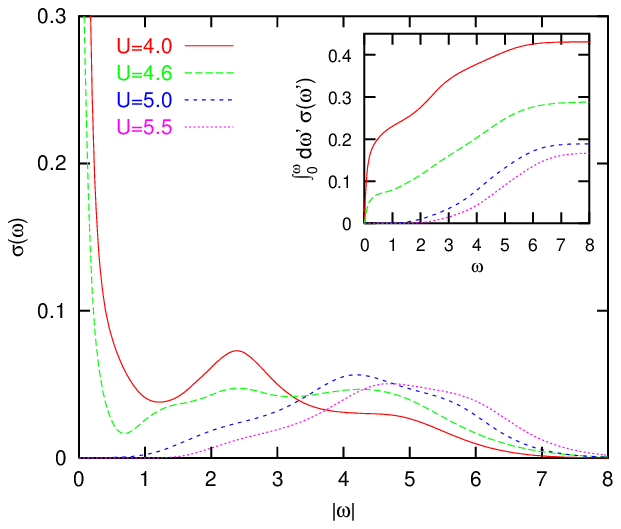}
\caption{Numerical results for the half-filled Hubbard model with
  semi-elliptic DOS for $d\to\infty$ in the paramagnetic phase at
  $T=0.05$.  {\bf a)}~Local spectral function $A(\omega)$ obtained
  from QMC (using a discretization $\Delta\tau=0.1$) and MEM.  {\bf
    b)}~Optical conductivity $\sigma(\omega)$ for the isotropic ``redefined Bethe
  lattice''. The inset shows the partial $f$-sum
  $\int_0^{\omega}\!\mathrm{d}\omega' \sigma(\omega')$.}
\label{fig:mit_spect_opt}
\end{figure}

The local spectral functions for $T=0.05$, i.e., slightly below the
critical temperature $T^*\approx 0.055$, are shown in
Fig.~\ref{fig:mit_spect_opt}a as obtained from QMC/MEM \cite{Bluemer02a}.  In
the metallic phase, the spectral density at the Fermi level
($\omega=0$) is approximately pinned at the noninteracting value
$\rho(0)=1/\pi\approx 0.32$ for $U\lesssim 4.4$. The quasiparticle
weight decreases drastically and a shoulder develops for $U\gtrsim4.6$
before a gap opens for $U\gtrsim 4.8$. An application of (\ref{eq:opt_dmft})
to these spectra for the isotropic model characterized by 
(\ref{eq:vksq_bethe_general}) yields the estimates for the 
optical conductivity $\sigma(\omega)$ shown in Fig.~\ref{fig:mit_spect_opt}b.
A low-frequency Drude peak (of Lorentzian form) and a mid-infrared peak
at $\omega\approx U/2$ are present in the metallic phase and decay
towards the metal-insulator transition at $U\approx 4.7$. For large
$U$, the optical spectral weight concentrates in incoherent peaks
at $\omega\approx U$. The inset of Fig.~\ref{fig:mit_spect_opt}b shows the
partial optical $f$-sums.  As expected,
both the contribution of the Drude peak and the total $f$-sum decrease
for increasing $U$.

\begin{figure}
a)\hspace{0.48\textwidth} b)\vspace{-2ex}

  \includegraphics[width=0.49\textwidth,clip=true]{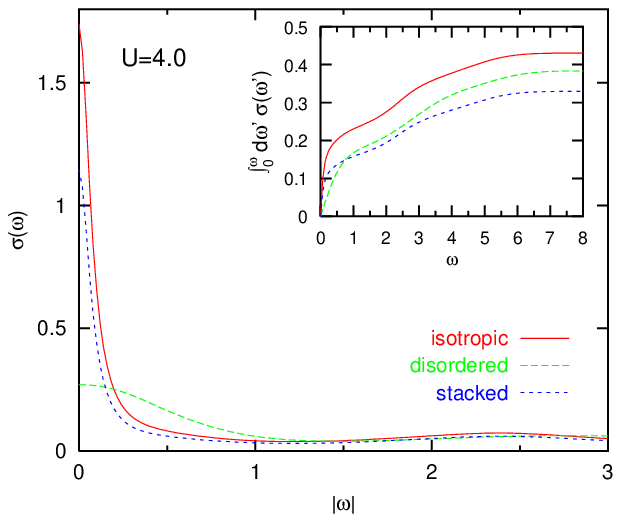}
 \hfill \includegraphics[width=0.49\textwidth,clip=true]{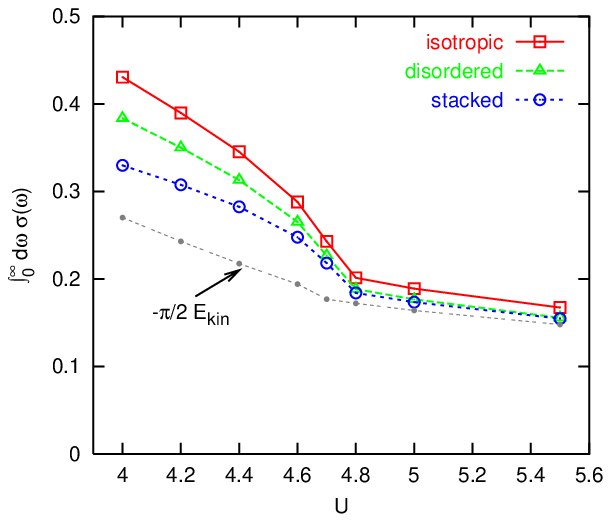}
\caption{{\bf a)} Optical conductivity $\sigma(\omega)$ for $T=0.05$ and $U=4.0$ for
the new isotropic model with semi-elliptic DOS in comparison with disordered
and stacked models consistent with the same DOS. 
{\bf b)} Optical $f$-sum for $T=0.05$.}
\label{fig:opt_fsum}
\end{figure}

Figure \ref{fig:opt_fsum}a shows the impact of the definition for
$\sigma(\omega)$ on the results for $U=4.0$. 
Our isotropic model yields by far the largest contributions at
small $\omega$. 
While the stacked model leads to otherwise similar
results, the low-frequency form of $\sigma(\omega)$ is qualitatively different
in the disordered case (where a Drude peak is absent even for $U\to 0$).
As seen in Fig.~\ref{fig:opt_fsum}b, 
the $f$-sum is generally larger for the isotropic than for the stacked
Bethe lattice, in particular in the metallic phase (while the
disordered case is in-between). These differences can be attributed to
the enhanced squared Fermi velocity in the isotropic model: The
enhancement is largest near the Fermi surface, at $\epsilon=0$ by a
factor of $\pi/2\approx 1.57$. A corresponding increase is expected of the 
Drude peak for small enough $U$ and $T$ when transport is dominated by
states with $\epsilon\approx 0$. Since energy eigenstates spread
out in momentum space at large $U$, the enhancement reduces (in general) to 
the integral $\int_{-\infty}^\infty \!\mathrm{d}\epsilon\,
\tilde{\rho}(\epsilon)=\sum_{D=1}^\infty D\, {t_D^*}^2$ which evaluates
here to 1.05406. 

In all three cases, the proportionality (\ref{eq:fsum_ekin}) of the
$f$-sum to the kinetic energy as characteristic for the hc lattice is
clearly violated. This is true even for the stacked case (which is
otherwise similar to the hc case): while the $f$-sum is here
proportional to the contribution to the kinetic energy associated with
hopping in current direction \cite{Uhrig93b}, $\int_0^\infty
\!\mathrm{d}\omega\,\sigma_{xx}(\omega)=-\sigma_0 \langle \hat{T}_x\rangle/4$, this
contribution (which is negligible in the limit $Z\to \infty$) is not
proportional to the total kinetic energy in this anisotropic case.
A more relevant sum rule is derived from (\ref{eq:fsum}):
$\int_0^\infty \!\mathrm{d}\omega\,\sigma_{xx}(\omega)=\frac{t^2 a^2 \sigma_0}{4}
\langle\frac{-\epsilon}{4-\epsilon^2}\rangle$.

\section{Conclusion}

We have presented a new general method for constructing regular
lattice models with hypercubic (hc) symmetry, i.e. isotropic optical
transport properties, in large dimensions.  Previously, calculations
of the optical conductivity $\sigma(\omega)$ of the Hubbard model in the limit
$d\to\infty$ had been restricted to the hypercubic lattice (using NCA
\cite{Pruschke93a} or QMC \cite{Jarrell95b}) or have ignored the
lattice dependence: Applying the hc formalism to  
the Bethe semi-elliptic DOS, Rozenberg et.~al \cite{Rozenberg95a}
overlooked violations of the hc $f$-sum rule. 
All previous approaches specific to the Bethe DOS
were associated with anisotropic or incoherent transport; the most
interesting of these \cite{Chung98a,Chattopadhyay00a} 
have not yet been linked rigorously to microscopic models. 

Our method yields the first derivation for $\sigma(\omega)$
consistent with a semi-elliptic DOS that implies isotropic transport
which is fully coherent in the noninteracting limit. This reinterpretation
of the ``Bethe lattice'' (in the DMFT sense) as an isotropic, regular and
clean lattice and the demonstration
that the associated transport properties are robust (with respect to
finite dimensionality or hopping range) removes, finally, the pathologies
previously associated with the DMFT treatment of transport in connection
with non-Gaussian DOSs. At essentially no additional cost, the method
can also be used for computing properties such as transverse conductivities 
and thermopower; these vanish, however, in the particle-hole symmetric
case considered in this paper. Our numerical results have shown that
the precise definition of $\sigma(\omega)$ does matter, in particular within the
metallic phase where transport is potentially most coherent.
We have also found a general DMFT expression for the $f$-sum rule as
well as a form specific to our new approach.

\begin{acknowledgements}
We gratefully acknowledge useful discussions with J.~Freericks, D.~Logan, 
A.~Millis, and D.~Vollhardt.
\end{acknowledgements}

\end{article}

\end{document}